\documentclass[final,authoryear,5p,times,twocolumn]{elsarticle}

\usepackage{pgfmath}
\usepackage{etoolbox} % true false 
\providetoggle{thesis}\settoggle{thesis}{false}
\providetoggle{beamer}\settoggle{beamer}{false}
\providetoggle{final}\settoggle{final}{true} % use smaller plots etc
\usepackage{lmodern}        % Use an extension of the original Computer Modern font to minimize the use of bitmapped letters.
\usepackage{courier}
\usepackage{listings}
\usepackage[T1]{fontenc}    % Determines font encoding of the output. Font packages have to be included before this line.
\usepackage[utf8]{inputenc} % Determines encoding of the input. All input files have to use UTF8 encoding.
\usepackage{graphicx}
\usepackage{xspace}

\usepackage{amsmath}    % Extended typesetting of mathematical expression.
\usepackage{amssymb}    % Provides a multitude of mathematical symbols.
\usepackage{mathtools}  % Further extensions of mathematical typesetting.
\usepackage{microtype}  % Small-scale typographic enhancements.
\usepackage[inline]{enumitem} % User control over the layout of lists (itemize, enumerate, description).
\usepackage{multirow}   % Allows table elements to span several rows.
\usepackage{booktabs}   % Improves the type settings of tables.
\usepackage{subcaption} % Allows the use of subfigures and enables their referencing.
\usepackage[usenames,dvipsnames,table]{xcolor} % Allows the definition and use of colors. This package has to be included before tikz.
\usepackage{nag}       % Issues warnings when best practices in writing LaTeX documents are violated.
\usepackage{todonotes} % Provides tooltip-like todo notes.
\usepackage{hyperref}  % Enables cross linking in the electronic document version. This package has to be included second to last.
\usepackage[acronym,toc]{glossaries} % Enables the generation of glossaries and lists fo acronyms. This package has to be included last.
\usepackage{marvosym}
\usepackage{nicefrac}
\usepackage{bm}
\usepackage[english]{babel}
\usepackage[]{cleveref}
\usepackage{soul}
\usepackage{soulutf8}
\usepackage{blkarray}
\usepackage{floatpag}
\usepackage{url}
\usepackage{pgf-pie} % ugly piechart package
\soulregister\cite7
\soulregister\ref7
\soulregister\cref7
\soulregister\pageref7
\soulregister\citeyear7

\usepackage{tikz}
\usepackage{pgfplots}
\usepackage{float}
\iftoggle{beamer}{}{
    \usepackage{ntheorem}

}

 \def\nBanks{796\space}
 
 \def\NWSize{\iftoggle{final}{5000}{500}c}
\usetikzlibrary{shapes,backgrounds,calc}
\usetikzlibrary{arrows.meta}
\usetikzlibrary{positioning,graphs}

\definecolor{uvred}{RGB}{255,31,0}
\definecolor{uvblue}{RGB}{0,105,170}

\definecolor{Red}{rgb}{0.89, 0.0, 0.13}
\definecolor{Green}{rgb}{0.01, 0.75, 0.24}
\definecolor{gv}{rgb}{0.61, 0.85, 0.24}
\definecolor{bv}{rgb}{0.26, 0.57, 0.66}
\colorlet{uvblue}{uvblue}
\colorlet{uvbluedark}{black!40!uvblue}
\colorlet{uvbluelight}{white!40!uvblue}

\pgfplotsset{
clickable coords/.style={}, %% comment this line out if coords should be clickable
plotl/.style={uvbluelight,no marks,ultra thick,samples=500},
plott/.style={uvblue,no marks,ultra thick,samples=500},
plotd/.style={uvbluedark,no marks,ultra thick,samples=500},
plotA/.style={uvblue!40!white,ultra thick,mark=otimes*},
plotB/.style={uvblue!70!white,ultra thick,mark=triangle*},
plotC/.style={uvblue!90!white,ultra thick,mark=pentagon*},
plotD/.style={uvblue!70!black,ultra thick,mark=diamond*},
plotE/.style={uvblue!50!black,ultra thick,mark=square*},
plotF/.style={uvblue!20!black,ultra thick,mark=otimes},
LplotA/.style={white!10!uvbluelight,ultra thick,},
LplotB/.style={uvbluelight,ultra thick,},
LplotC/.style={uvblue,ultra thick},
LplotD/.style={uvbluedark,ultra thick},
LplotE/.style={black!40!uvbluedark,ultra thick},
LplotF/.style={black!60!uvbluedark,ultra thick},
asdfas/.style={}
}

\pgfplotsset{
    discard if not/.style 2 args={
        x filter/.code={
            \edef\tempa{\thisrow{#1}}
            \edef\tempb{#2}
            \ifx\tempa\tempb
            \else
                \def\pgfmathresult{inf}
            \fi
        }
    }
}

\pgfplotstableset{
    discard if not/.style 2 args={
        row predicate/.code={
            \def\pgfplotstable@loc@TMPd{\pgfplotstablegetelem{##1}{#1}\of}
            \expandafter\pgfplotstable@loc@TMPd\pgfplotstablename
            \edef\tempa{\pgfplotsretval}
            \edef\tempb{#2}
            \ifx\tempa\tempb
            \else
                \pgfplotstableuserowfalse
            \fi
        }
    }
}

\pgfplotsset{
    /pgfplots/flexible xticklabels from table/.code n args={3}{%
        \pgfplotstableread[#3]{#1}\coordinate@table
        \pgfplotstablegetcolumn{#2}\of{\coordinate@table}\to\pgfplots@xticklabels
        \let\pgfplots@xticklabel=\pgfplots@user@ticklabel@list@x
    }
}

\colorlet{hlblue}{uvred!30}
\sethlcolor{hlblue}

\usetikzlibrary{backgrounds}

\usepgfplotslibrary{colormaps}
\pgfplotsset{colormap name=viridis}%
\tikzset{
    ellEV/.style={
        color of colormap={200 + #1*1600}, % offset * 2 + scale <= 2000
        draw=.!80!black,
        fill=.!70!white,
        text = black},
    degCen/.style={
        color of colormap={200 + #1 * 800/7}, % offset +  + scale <= 1000
        draw=.!80!black,
        fill=.!70!white,
        text = black
    },
    degCen2/.style={
        color of colormap={300 + #1 * 700/5}, % offset + scale <= 1000
        draw=.!80!black,
        fill=.!70!white,
        text = black
    },
    drMap/.style={
        color of colormap={250 + #1 * 750}, % offset + scale <= 1000
        draw=.!80!black,
        fill=.!70!white,
        text = black
    },
    bankCount/.style={ % between 0 and 6
    ,    color of colormap={#1 * 1000/6} % offset + scale <= 1000
    ,    draw=.!60!black
    ,    fill=.!100!white
    ,    text = black
	},
    onaceColor/.style={ % between 0 and 10
    ,    color of colormap={#1 * 111} % offset + scale <= 1000
    ,    draw=.!60!black
    ,    fill=.!100!white
    ,    text = black
    },
    bankColor/.style={
    ,    color of colormap={850}
    ,    draw=.!70!black
    ,    fill=.!100!white
    ,    text = black
    },
    companyColor/.style={
    ,   color of colormap={400}
    ,   draw=.!70!black
    ,   fill=.!100!white
    ,   text = black
    },
    bankColorLine/.style={
    ,    color of colormap={850}
    ,    draw=.!100!white
    ,    fill=.!100!white
    ,    text = black
    },
    companyColorLine/.style={
    ,   color of colormap={400}
    ,   draw=.!100!white
    ,   fill=.!100!white
    ,   text = black
	},
    depositColor/.style={
    ,   color of colormap={1000}
    ,   draw=.!70!black
    ,   fill=.!100!white
    ,   text = black
	},
	adjMatrix/.style={
    ,   color of colormap={#1 * 825/5 + 125}
    ,   draw=.!100!black
    ,   fill= none
	},
	diffGlg/.style={ % between 0 and 10
    ,    color of colormap={#1 * 111} % offset + scale <= 1000
    },
}

\pgfplotscreateplotcyclelist{connCountList}{
	{bankCount=6},
	{bankCount=5},
	{bankCount=4},
	{bankCount=3},
	{bankCount=2},
	{bankCount=1},
	{bankCount=0},
}
\pgfplotscreateplotcyclelist{onaceColors}{
	{onaceColor=0},
	{onaceColor=9},
	{onaceColor=4},
	{onaceColor=5},
	{onaceColor=3},
	{onaceColor=2},
	{onaceColor=7},
	{onaceColor=1},
	{onaceColor=8},
	{onaceColor=6},
}
\pgfplotscreateplotcyclelist{compBankList}{
	{companyColor},
	{bankColor},
}

\makeatletter
\pgfutil@namelet{pgf@sh@fbg@circle split@original}{pgf@sh@fbg@circle split}%
\newif\ifpgfshapecirclesplitdrawsplits
\pgfshapecirclesplitdrawsplitstrue
\pgfkeys{%
  /pgf/circle split draw splits/.is if=pgfshapecirclesplitdrawsplits
}
\def\pgf@sm@shape@name{circle split}
\pgf@sh@beforebgpath{%
  \ifpgfshapecirclesplitdrawsplits%
    \csname pgf@sh@fbg@circle split@original\endcsname%
  \fi
}

\tikzset{circle split part fill/.style  args={#1,#2,#3}{%
 alias=tmp@name, % Jake's idea !!
  postaction={%
    insert path={
     \pgfextra{% 
     \pgfpointdiff{\pgfpointanchor{\pgf@node@name}{center}}%
                  {\pgfpointanchor{\pgf@node@name}{east}}%            
     \pgfmathsetmacro\insiderad{\pgf@x}
      \fill[#3=#1]
	  		(\pgf@node@name.base)
			([xshift=-\pgflinewidth]\pgf@node@name.east)
            arc (0:180:\insiderad-\pgflinewidth);
      \fill[#3=#2]
            (\pgf@node@name.base)
            ([xshift=\pgflinewidth]\pgf@node@name.west)
            arc (180:360:\insiderad-\pgflinewidth)--cycle;
         }}}}}  
\makeatother  

\tikzset{
    bank/.style={
        ,   regular polygon
        ,   regular polygon sides=4
        ,   bankColor
    },
    company/.style={
        ,   circle
        ,   companyColor
    }
}


\usepackage{afterpage}
\newcommand\atevenpage[1]{%
  \afterpage{\clearpage% be sure, that there are no pending floats
    \ifodd\value{page}% still a odd page
      \atevenpage{#1}%
    \else
      #1%
    \fi
  }%
}
\newcommand\atoddpage[1]{%
  \afterpage{\clearpage% be sure, that there are no pending floats
    \ifodd\value{page}% still a odd page
      #1%
    \else
      \atevenpage{#1}%
    \fi
  }%
}

\DeclareRobustCommand{\capNode}[1]{\tikz[anchor=base,baseline=-.125cm]{\node[#1] (l) at (0,0) {};}\xspace}
\DeclareRobustCommand{\capEdge}[1]{\tikz[scale=0.8,anchor=base,baseline=-.125cm]{\draw[#1, -{Latex[length=1.5mm, width=1mm]}] (0,-0.25)--(0.5,0.25);}\xspace}

\newcommand{\plotref}[1]{{[\,\ref{#1}\,]}\xspace}

\newcommand{\cbnScale}{1.7}

\pgfplotsset{
        colormap={RdYlGn}{
            rgb255=(0, 104, 55),
            rgb255=(255, 255, 190),
            rgb255=(166, 0, 38),
		},
        colormap={ugly}{
            rgb255=(255, 0, 0),
            rgb255=(255, 255, 0),
            rgb255=(0, 255, 0),
		},
		colormap={debtrank}{
            rgb255=(0, 255, 0),
            rgb255=(255, 255, 0),
            rgb255=(255, 0, 0),
		},
		colormap/viridis
    }
\crefformat{appendix}{#2#1#3}
\hypersetup{draft}
\DeclareRobustCommand{\capNode}[1]{\tikz[scale=0.9,anchor=base,baseline=-.125cm]{\node[#1] (l) at (0,0) {};}\xspace}

\begin{document} 
\begin{frontmatter}
	
	\title{Identifying systemically important companies in the entire liability network of a small open economy}
	
	\author[iiasa,hub]{Sebastian Poledna} \ead{poledna@iiasa.ac.at}
	\author[hub,cosy]{Abraham Hinteregger} \ead{oerpli@outlook.com}
	\author[cosy,hub,sfi,iiasa]{Stefan Thurner\corref{cor}} \ead{stefan.thurner@meduniwien.ac.at}
	\cortext[cor]{Corresponding author}
	
	\address[iiasa]{IIASA, Schlossplatz 1, A-2361 Laxenburg, Austria}
	\address[hub]{Complexity Science Hub Vienna, Josefst{\"a}dter Stra{\ss}e 39, 1080 Vienna, Austria} 
	\address[cosy]{Section for Science of Complex Systems, Medical University of Vienna, Spitalgasse 23, A-1090, Austria} 
	\address[sfi]{Santa Fe Institute, 1399 Hyde Park Road, Santa Fe, NM 87501, USA} 
	
	\begin{abstract}
		To a large extent, the systemic importance of financial institutions is related to the topology of financial liability networks. In this work we reconstruct and analyze the -- to our knowledge -- largest financial network that has been studied up to now. This financial liability network consists of 51,980 firms and \nBanks banks. It represents $80.2\%$ of total liabilities towards banks by firms and all interbank liabilities from the entire Austrian banking system. We find that firms contribute to systemic risk in similar ways as banks do. In particular, we identify several medium-sized banks and firms with total assets below 1 bln. EUR that are systemically important in the entire financial network. We show that the notion of systemically important financial institutions (SIFIs) or global and domestic systemically important banks (G-SIBs or D-SIBs) can be straightforwardly extended to firms. We find that firms introduce slightly more systemic risk than banks. In Austria in 2008, the total systemic risk of the interbank network amounts to only $29\%$ of the total systemic risk of the entire financial network, consisting of firms and banks.
	\end{abstract}
	
	
\end{frontmatter}

\section{Introduction}
The financial crisis of 2007–2008 was sparked by the collapse of a relatively small -- now famous -- investment bank and propagated through the financial system, bringing the world financial system to the brink of collapse. Through interrelationships between the financial and the real economy, the financial crisis spread quickly and was followed by a global economic downturn, the Great Recession. Potentially, also the opposite could happen: a financial crisis could originate in the real economy and spread to the financial system, making the study of interrelationships between the financial and the real economy more important than ever.

In response to the financial crisis, the Basel III framework recognizes systemically important financial institutions (SIFIs) and, in particular, global and domestic systemically important banks (G-SIBs or D-SIBs) and recommends increased capital requirements for them -- the so called ``SIFI surcharges'' \citep{BIS:2010aa}. In this context several network-based measures that identify systemically important financial institutions have been proposed and applied recently \citep{Battiston:2012aa,Markose:2012aa,Billio:2012aa,Thurner:2013aa,Poledna:2016ab,Leduc:2016ab,Poledna:2017aa}. These approaches bear the notion of the {\em systemic importance} of a financial institution within a financial network and rely on {\em network centrality} or on closely related measures.

These network-based approaches typically work well in small financial networks like banking networks with a relatively small number of financial institutions, usually less than a thousand. A serious disadvantage of centrality measures is, however, that the corresponding value for a particular institution has no clear interpretation as a measure for expected losses. A solution that solves this problem is the so-called ``DebtRank'', a recursive method, suggested by \citet{Battiston:2012aa}, which quantifies the systemic importance of financial institutions in terms of losses that they would contribute to the total loss in case of a default. Since data on financial networks is hard to obtain outside central banks and is typically only available for small banking networks, there have also been several attempts to quantify systemic importance of institutions without the explicit knowledge of the underlying networks \citep{Adrian:2011aa,Acharya:2010aa,Brownlees:2012aa,Huang:2012aa}.

These developments have spurred research on financial networks. Driven by data availability, research on financial networks has mainly focused on default contagion: mostly, on direct lending networks between financial institutions \citep{Upper:2002aa,Boss:2004aa,Boss:2005aa,Soramaki:2007aa,Iori:2008aa,Cajueiro:2009aa,Bech:2010aa,Fricke:2014aa,Iori:2014aa}, but also on the network of derivative exposures \citep{Markose:2012aa,Markose:2012ab}. Research on financial \emph{multi-layer} networks that considers multiple channels of contagion, has only appeared recently. \citet{Poledna:2015aa} and \citet{Leon:2014aa} study the interactions of financial institutions on different financial markets in Mexico and Colombia, respectively.

To our knowledge, there exist only a few works that empirically study the interrelationship between the financial and the real economy. These studies focus on Japan and are mainly concerned with the topology of credit networks between banks and large firms \citep{Fujiwara:2009aa,De-Masi:2012aa,Aoyama:2014aa,Marotta:2015aa}. Additionally, \citet{De-Masi:2011aa} and \cite{Miranda:2013aa} study credit networks in Italy and Brazil, and \citet{Lux:2016aa} develops a theoretical model of a bipartite credit network between banks and the non-bank corporate sector. \citet{De-Masi:2011aa} and \citet{De-Masi:2012aa} use network analysis to study the credit networks in Italy and Japan and \citet{Fujiwara:2009aa} and \citet{Marotta:2015aa} investigate the evolution of the network structure in Japan. \citet{Marotta:2015aa} further perform community detection, identifying communities composed of both banks and firms. 

\citet{Miranda:2013aa} and \citet{Aoyama:2014aa} make the first attempt to empirically study systemic risk in credit networks in Japan and Brazil. \cite{Aoyama:2014aa} uses DebtRank to study risk propagation from banks to firms with a dataset provided by Nikkei Inc. that contains approximately 2,000 firms and 200 banks in Japan, but does not include interbank data. \citet{Miranda:2013aa} perform the first study that includes interbank and firm loans. However, the used dataset is relatively small and contains only about 50 banks and 351 firms in Brazil.

In this work we extend the existing literature by reconstructing and analyzing a large financial network that not only includes \emph{all} interbank liabilities but also nearly all liabilities and deposits between banks and firms. We do this by combing datasets that contain annual financial statements of nearly all firms and banks in Austria (approximately $170,000$ firms and close to $1,000$ banks) with anonymized interbank liabilities from the Austrian banking system.
The reconstruction of this large financial network of firms and banks allows us to identify systemically important firms by employing DebtRank to financial networks as in \citet{Thurner:2013aa}. We estimate the share of systemic risk introduced by firms and we compare the systemic risk of the interbank network with the systemic risk of the entire financial network consisting of firms and banks. We show that the notion of systemically important financial institutions (SIFIs) or global and domestic systemically important banks (G-SIBs or D-SIBs) can be straightforwardly extended to firms. 

The paper is structured as follows. \Cref{sec:data} provides an overview of the datasets used in this study. In \cref{sec:construction} we explain the methodology to reconstruct the entire financial network. In \cref{sec:results,sec:statistics} we present the results by first presenting classical network statistics of the entire financial network, followed by an analysis of systemic importance of firms and banks. Finally, \cref{sec:discussion} discusses the results and provides conclusions.

\section{Data}\label{sec:data}
The data used for this analysis consists of two main parts, annual financial statements of nearly all firms and banks in Austria and anonymized interbank liabilities from the Austrian banking system.
Financial statements of firms are taken from the SABINA database\footnote{The SABINA database is provided by Bureau van Dijk, see \href{https://www.bvdinfo.com/en-us/our-products/company-information/national-products/sabina}{https://www.bvdinfo.com/en-us/our-products/company-information/national-products/sabina}.}, which provides information on about $170,000$ firms in Austria. This database contains company financials in a detailed format with up to 10 years of history, and additionally, data on shareholders and subsidiaries, activity codes and trade descriptions, and stock data for listed companies. The database includes bank-firm relationships and allows us to identify which firm is a customer of which bank. 

Financial statements of banks are made publicly available by the Austrian Central Bank (OeNB)\footnote{\href{https://www.oenb.at/jahresabschlusski/jahresabschlusski}{https://www.oenb.at/jahresabschlusski/jahresabschlusski}}. 
Interbank data provided by the OeNB contains fully anonymized and linearly transformed interbank liabilities from the entire Austrian banking system over 12 consecutive quarters from 2006-2008. The dataset additionally includes total assets, total liabilities, assets due from banks, liabilities due to banks, and liquid assets (without interbank assets/liabilities) for all banks, again in anonymized form. 

\begin{figure}
	\centering
	% !TeX root = ../MSc.tex
\begin{tikzpicture}[]
\def\file{./data/histLiabTransposed.csv}
 \pgfplotstableread[col sep=tab,trim cells]{\file}\table
%\pgfplotstableread[col sep=tab,trim cells]{\fileB}\tableB
%\pgfplotstableread{./data/returnhistF.csv}\loadedtableF
\begin{axis}[width=\iftoggle{thesis}{1.0}{0.53}\textwidth
,	height=350pt
,	xbar stacked
,	xlabel={Sum of liabilities (EUR)}
,	ylabel={Liability types ordered by their total volume}
,	ylabel style={yshift=-25pt}
%,	border={1pt 1pt 1pt 0pt}
%,	ytick=data
,	ytick = \empty
%,	xtick = {1E11,...,5E11}
%,	yticklabels from table={\table}{name}
,	every node near coord/.append style={rotate=0, anchor=west,font=\footnotesize  }
%,	scaled x ticks = false
,	xmajorgrids,
%,	axis x line*=bottom
,	axis y line*=left
%,	hide y axis
,	bar width= 8pt
,	xmin = 0
,	xmax = 5.5E11
,	ymax = 0
,	ymin =-31
,	x tick style={opacity=0}
,	major grid style={thin,dashed,black!20}
,	legend pos=south east
,	legend style={fill=white, fill opacity=0.9, draw opacity=1,text opacity=1, draw = none}
,	legend style={
		,	fill=white
		,	draw opacity=1
		,	text opacity=1
		% ,	draw=none
		% ,	legend columns=-1
		% % ,	fill opacity=0.9
		% ,	column sep=0.28ex
		% ,	yshift=8pt
		% ,	xshift=8pt
	}
% ,	legend image post style={xscale=1.65,yscale=1.8}
,	legend image post style={xscale=1.4,yscale=1.4}
,	legend cell align=left
,	cycle list name=connCountList
,	]
\pgfplotsinvokeforeach{0,1,...,5}
{
\addplot+[]
	table[trim cells,y expr= -\coordindex+0.5, x =c#1] from \file;\label{plt:liab#1}\addlegendentry{{#1 Bank\ifthenelse{#1 = 1}{}{s}}}
}
\addplot+[nodes near coords,nodes near coords align={vertical},
	point meta=explicit symbolic] 
	table[trim cells,y expr= -\coordindex+0.5, x =c6, meta = name] from \file;
	\addlegendentry{{6 Banks}} 
\end{axis}
% white rectangles to hide dotted lines
% \draw[fill=white,draw=none](6.5,1.1) rectangle(16,5);
% \draw[fill=white,draw=none](8,0.65) rectangle(13,0.2);
% \node[] at (11,2.5) {\input{./img/bankCountHistSmall.tex}};% small plot in plot

\end{tikzpicture}
	\caption{Types of different liabilities aggregated over the number of banks associated with each firm. Data is obtained from the balance sheets filed by each firm in the year 2008. Different liabilities are sorted according to total volume in each type.}\label{fig:liabhist}
\end{figure}
There are $106,919$ firms and \nBanks banks that provided a financial statement in the calendar year 2008. 
\Cref{fig:liabhist} shows a stacked bar chart of the different liabilities found in the balance sheets filed in the calendar year 2008, aggregated by the number of banks associated with each firm.
\begin{figure}
	\centering
	% !TeX root = ../MSc.tex
\begin{tikzpicture}[]

\def\file{./data/bankCountHist.csv}
 \pgfplotstableread[col sep=tab,trim cells]{\file}\table
% \pgfplotstableread{./data/returnhistF.csv}\loadedtableF
  \begin{axis}[width=\iftoggle{thesis}{0.7}{0.4}\textwidth
,	height=115pt
,	ybar
,	xlabel={Number of bank connections}
,	ylabel={Number of firms}
,	every node near coord/.append style={rotate=0, anchor=south,font=\footnotesize}
%,	xticklabels from table={\loadedtable}{label},
%,	flexible xticklabels from table={\file}{label}{col sep=tab}
,	xticklabel style={text height=1.5ex} % To make sure the text labels are nicely aligned
,	xtick={0,1,2,3,4,5,6}
,	ytick={0,25000,50000}
,	scaled x ticks = false
,	ymin = 0,	ymax =55000
%,	xmin = 0, xmax = 29
,	ymajorgrids,
,	axis y line*=left
,	bar width= 15pt
,	ybar =-15pt
%,	hide x axis
,	axis x line*=bottom,
,	major grid style={thin,dashed,black!20}
,	scaled y ticks = false
,	ylabel style={yshift=\iftoggle{thesis}{15pt}{5pt}}
,	xtick style={opacity=0}
,	ytick style={opacity=0}
,	y tick label style=
		{/pgf/number format/fixed
		,/pgf/number format/1000 sep = \thinspace % replace comma as 1000 separator 
		}
,	cycle list name=connCountList,
%,	legend pos=south west
%,	legend style={fill=white, fill opacity=0.9, draw opacity=1,text opacity=1}
%,	legend cell align=left
 ]
  \pgfplotsinvokeforeach{0,1,2,3,4,5,6}{
	\addplot+[nodes near coords,nodes near coords align={vertical},
		point meta=explicit symbolic]	table[trim cells,x expr=#1, y index =#1, meta index= #1] from \file;
    }
%	\addlegendentry{\#Companies }
  \end{axis}
\end{tikzpicture}
	\caption{Number of banks associated with each firm.}\label{fig:bankhist}
\end{figure}
In \cref{fig:bankhist} the number of banks associated with each firm is shown. Approximately $48.6\%$ of the firms representing about $80.2\%$ of total liabilities towards banks can be associated with one or more banks. Firms that can not be associated with a bank are excluded from the analysis. Since it is not compulsory for small firms to provide an exact breakdown of  different liabilities, we estimate liabilities towards banks with the average ratio of firms in the same line of business, as indicated by their OeNACE code~\cite{wirtschaftskammer_osterreich_onace_2008}.  We now reconstruct the liability network of \nBanks banks and 51,980 firms representing $80.2\%$ of total liabilities towards banks by firms and all interbank liabilities. 

\section{Reconstruction of the liability network} \label{sec:construction}
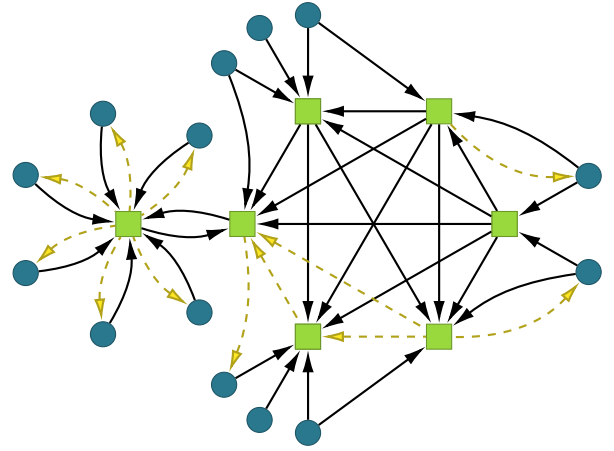
\begin{figure}
	\centering
	\renewcommand{\cbnScale}{1.5}
	% !TeX root = ../MSc.tex
\begin{tikzpicture}[scale=\cbnScale, {Latex[length=3mm, width=1.5mm]}-, deposit/.style={depositColor, dashed}] 
\coordinate (C1) at (2,2);

% Flowerlike thingy on left 
\node[bank] (X) at (C1){}; % center 1 finished
\foreach \a in {1,...,6}{
  \node[company] (C) at ($(X)+(360/7*\a:1)$){};
  \path[thick, deposit](C) edge [bend left=18] (X);
  \path[thick](X)  edge [bend left=18] (C);
}

% Specify center of hexagon of banks on right
\coordinate (M) at ($(C1)+(0:2.15)$);

\foreach \a in {0,...,5}{ % Hexagon
    \node[bank] (B\a) at ($(M)+(360/6*\a:1.15)$) {};
}

% Connect bank3 of hexagon with flower center
\path[thick](X)  edge [bend left=18] (B3);
\path[thick](B3)  edge [bend left=18] (X);

% Draw lines from  Bank a to Bank b iff a is smaller than b
\foreach \a in {0,...,5}{
    \foreach \b in {\a,...,5}{
        \ifthenelse{\a=\b}{}{
			\ifthenelse{\a > 2}{
%				\path[thick](B\b) edge[bend left=10] (B\a);
				\path[thick, deposit](B\a) edge[] (B\b);
			}{
				\path[thick](B\b) edge(B\a);
			}
        }
    }
}

% Add 3 companies to Bank 0,2,4
\foreach \a in {0,1,2}{
    \node[bank] (BX) at ($(M)+(240 +120*\a:1.15)$) {};
    \foreach \b in {0,...,2}{
        \ifthenelse{\a=\b \AND \a = 1}{}{
            \node[company] (C\a\b) at ($(BX)+(210+120*\a+30*\b:0.85)$) {};
            \draw[thick](BX) -- (C\a\b);
        }
    }
}

% Extra connections for three companies on far right
\path[thick](B1)  edge [bend left=15] (C12);
\path[thick, deposit](C12) edge [bend left=25] (B1);
\path[thick, deposit](C10) edge [bend left=25] (B5);
\path[thick](B5)  edge [bend left=15] (C10);

% Extra connections for three companies on top/bottom
\path[thick](B3) edge[bend right=15] (C22);
\path[thick](B1) edge[] (C20);
\path[thick](B5) edge[] (C02);
%\path[thick](B3) edge[bend left=15] (C00);
\path[thick, deposit](C00) edge[bend right=15] (B3);

%\iftoggle{thesis}{}{\node[opacity = 0] (push) at (2,-0.6){X};}
\end{tikzpicture} 
	\caption{Illustration of a network of banks \capNode{bank} and firms \capNode{company}. Connections are either loans \capEdge{thick} or deposits \capEdge{depositColor, dashed,thick}. Banks are connected to firms and to each other, whereas firms only interact with banks.}\label{fig:network:schema}
\end{figure}
We combined datasets from different sources to create a network that represents the liabilities and assets of the Austrian economy. The resulting bipartite network $G = (F,E)$ consists of two disjunct sets of nodes: banks $B$ and firms $C$, for which the following equations hold, 
\begin{align*}
	B\cup C &= F & B\cap C &= \emptyset\\
	|B| &= b & |C| &= c\ .
\end{align*}

The links either connect banks with other banks (interbank liabilities), banks with firms (deposits of firms at banks), or firms with banks (liabilities of firms).
The {weighted adjacency matrix} 
\begin{align}
	L_{n \times n } = \begin{pmatrix}
		BB_{b\times b}
		&
		BC_{b\times c}
		\\
		CB_{c\times b}
		&
		CC_{c\times c} = 0
	\end{pmatrix}, \qquad n = c+b\label{eq:liabmat}\,,
\end{align}
is called the \emph{liability matrix}, where each entry $L_{ij}$ indicates the liability that node $i$ (which is a bank if $i \leq b$ and a firm if $i > b$) has to node $j$ (the same applies here). It is partitioned into the following four parts:
\begin{enumerate}
	\item interbank network $BB$, connecting banks with banks,
	\item bank-firm network $BC$, containing information about deposits firms have at financial institutions, 
	\item firm-bank network $CB$, containing information about liabilities firms have to financial institutions,
	\item firm-firm network $CC$ with inter-firm liabilities, omitted in this work thus $CC_{xy} = 0$ for all $x,y \in [1,c]$. 
\end{enumerate}

The interbank network $BB$ is taken directly from the interbank dataset. Data on bank-firm relationships is used to establish an unweighted bipartite network between firms and banks. This bipartite network is used as a basis for the $BC$ and $CB$ adjacency matrices, which are (after assigning weights, see below) combined with $BB$ and padded with zeros to obtain the liability network $L$. To match the interbank network $BB$ with the bipartite bank-firm networks $BC$ and $CB$ the banks of both datasets were ranked according to total assets. The resulting tables were then joined with their rank as common column.

The weights of the bipartite liability network of firms and banks are assigned the following way:
\begin{itemize}
	\item For every firm $c$, we take the aggregated liabilities $L_c$ the firm has toward banks from the balance sheet. 
	\item We then take the set of aggregated loans (referred to as assets, or $A_i$ where $i$ is the index of a bank/firm) of all banks from their balance sheets and assign them to the entries of the vector $\ell$ in the following way: 
	\begin{equation*}
		\ell_b = \begin{cases}
			0 &\text{if $L_{cb} = 0$}\\
			A_b &\text{else}
		\end{cases}
	\end{equation*}
	\item We normalize the resulting vector with the L1 norm, 
	\begin{equation*}
		\hat{\ell }= \frac{\ell}{\sum_i |\ell_i|}\,.
	\end{equation*}
	\item We partition the aggregated liabilities with the distribution $\hat\ell$ to obtain the entries for the firm-bank network $L_{:c} = L_c\cdot \hat\ell\,$.
\end{itemize}
Note that we partition the liabilities of each firm to the banks to which it is connected to, according to the relative size of the banks.

\section{The liability network of Austria} \label{sec:statistics}
\begin{figure}
	\centering
	%\fbox
	\includegraphics[width=\columnwidth]{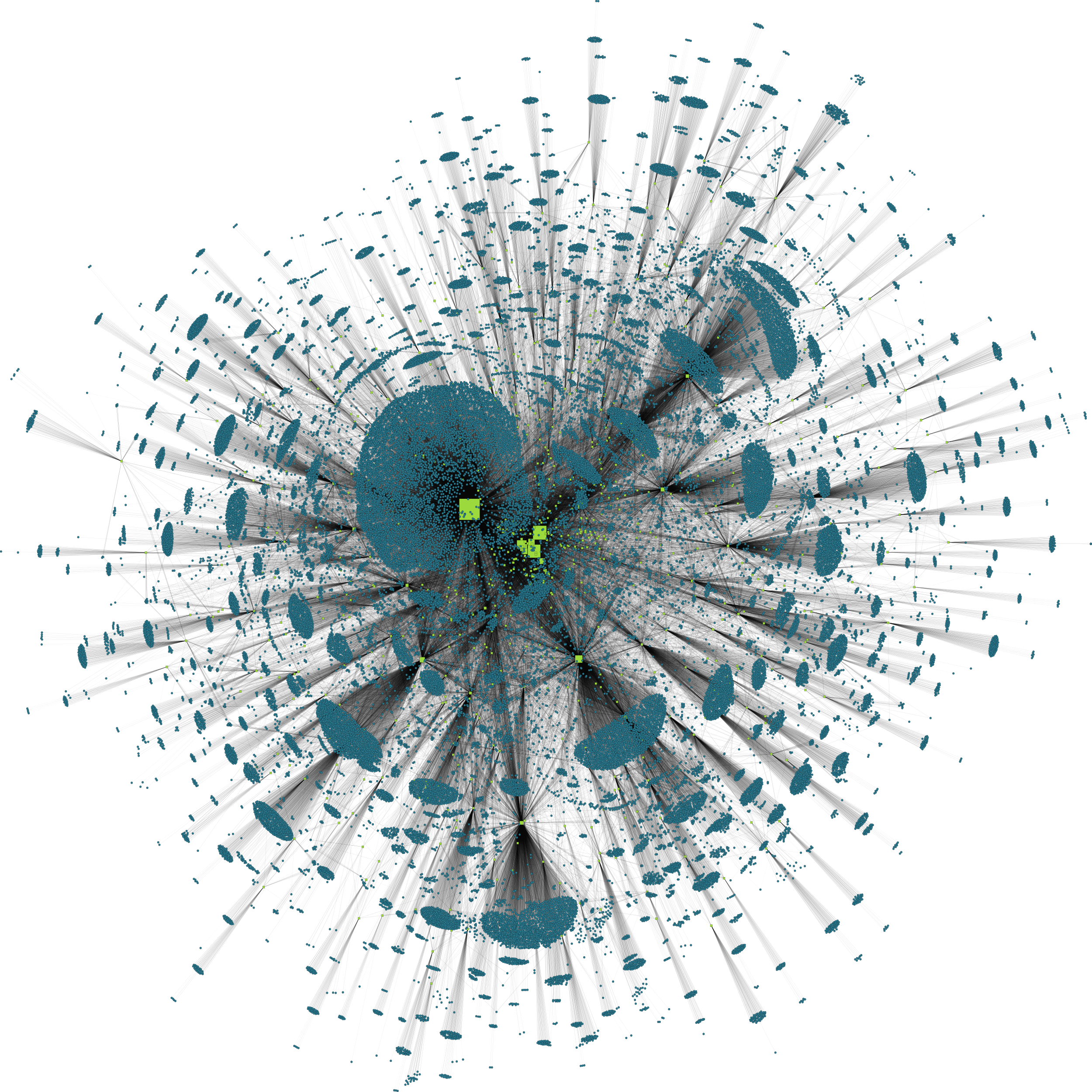}
	\caption{Reconstructed liability network of Austria with \nBanks bank nodes \capNode{bank} and 51,980 firm nodes \capNode{company} in 2008. The network represents approximately $80.2\%$ of total liabilities towards banks of firms and all interbank liabilities. Node size corresponds to the total assets held by each node.}
	\label{fig:graph:50k}
\end{figure}
We use empirical data (see \cref{sec:data}) to reconstruct the liability network of Austria, as outlined in \cref{sec:construction}. The resulting network with 52,980 nodes is visualized in \cref{fig:graph:50k} and represents approximately $80.2\%$ of total liabilities towards banks of firms and all interbank liabilities\footnote{We used the network layout algorithm from Hu Yifan~\cite{hu_efficient_2006} implementation in Gephi~\cite{bastian_gephi:_2009} to create the network illustrations in this work.}. Bank nodes are represented by squares and firms by circles. Node size corresponds to the total assets held by each node. For the following analysis we chose the subgraph induced by the set of all 796 banks in the Austrian banking system and the 5,000 firms with the highest liabilities. 
	
\begin{figure*}
	\centering
	% !TeX root = ../../MSc.tex
\begin{tikzpicture}
	\def\DRTX{./data/5000c/hists/BDeg.csv}
	\def\DRT{./data/5000c/hists/BDeg2.csv}
	\begin{axis}[
		,	width=0.95\textwidth
		,	height=170pt
		%,	axis on top
		,	scale only axis
		,	ymode=log
		%,	xmode=log
		,	ybar = 0pt% -3pt
		,	legend cell align = left
		,	bar width= 3.75pt
		%,	x tick label as interval=false
		%,	xtickten={-16,-14,-12,...,4}
		%,	yticklabel={}
		,	xtick pos=left
		,	ytick pos=left
		,	xmin=-40
		,	xmax=3000
		,	ymin=0.5
		,	ymax=1000
		%,	grid=none
		,	ymajorgrids
		,	log origin=infty
		,	cycle list name=compBankList
		,	ylabel = {Count}
		,	xlabel = {Degree of banks in the entire network}
		,	ylabel near ticks
		]
		\addplot+[,draw=none] table [x=left,y=in] {\DRT};
		\addplot+[,draw=none] table [x=left,y=out] {\DRT};
	\end{axis}
	\begin{axis}[
		,	width=0.79\textwidth
		,	xshift=2.6cm
		,	yshift=2.1cm
		,	height=105pt
		,				,	axis background/.style={fill=gray!02}
		,	scale only axis
		,	ymode=log
		%,	xmode=log
		,	ybar = 0pt% -3pt
		,	legend cell align = left
		,	legend style={draw=none}
		,	bar width= 1.75pt
		%,	x tick label as interval=false
		%,	xtickten={-16,-14,-12,...,4}
		%,	yticklabel={}
		,	xtick pos=left
		,	ytick pos=left
		,	xmin=-1
		,	xmax=100
		,	ymin=0.5
		%,	ymax=1
		%,	grid=none
		,	ymajorgrids
		,	log origin=infty
		,	cycle list name=compBankList
		,	ylabel = {Count}
		,	xlabel = {Degree of banks in the entire network}
		,	ylabel near ticks
		,	ylabel style={fill = white}
		]
		\addplot+[,draw=none] table [x=left,y=in] {\DRTX};\addlegendentry{Indegree}
		\addplot+[,draw=none] table [x=left,y=out] {\DRTX};\addlegendentry{Outdegree}
	\end{axis}
\end{tikzpicture}
	\caption{In- and out-degree of banks in the entire liability network $F$. Main plot has 60 uniform bins on the interval $[0,3000]$, the inset shows the distribution of degrees in the range $[0,100]$.}%
	\label{fig:deg:f:inout}
\end{figure*}
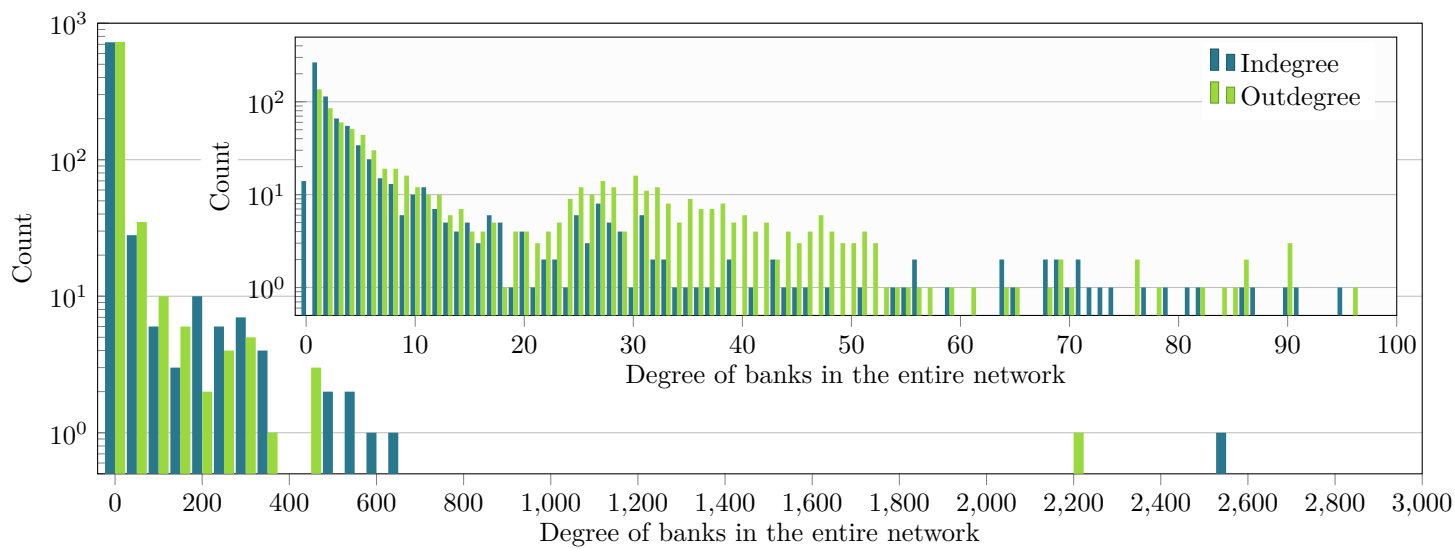
\begin{figure*}
	\centering
	% !TeX root = ../../MSc.tex
\begin{tikzpicture}
	\def\DRTX{./data/5000c/hists/IBDeg.csv}
	\def\DRT{./data/5000c/hists/IBDeg2.csv}
	\begin{axis}[
		,	width=0.95\textwidth
		,	height=170pt
		%,	axis on top
		,	scale only axis
		,	ymode=log
		%,	xmode=log
		,	ybar = 0pt% -3pt
		,	legend cell align = left
		,	bar width= 3.75pt
		%,	x tick label as interval=false
		%,	xtickten={-16,-14,-12,...,4}
		%,	yticklabel={}
		,	xtick pos=left
		,	ytick pos=left
		,	xmin=-5
		,	xmax=360
		,	ymin=0.5
		,	ymax=2000
		%,	grid=none
		,	ymajorgrids
		,	log origin=infty
		,	cycle list name=compBankList
		,	ylabel = {Count}
		,	xlabel = {Degree of banks in interbank network}
		,	ylabel near ticks
		]
		\addplot+[,draw=none] table [x=left,y=in] {\DRT};
		\addplot+[,draw=none] table [x=left,y=out] {\DRT};
	\end{axis}
	\begin{axis}[
		,	width=0.79\textwidth
		,	xshift=2.6cm
		,	yshift=2.6cm
		,	height=90pt
		,				,	axis background/.style={fill=gray!02}
		,	scale only axis
		,	ymode=log
		%,	xmode=log
		,	ybar = 0pt% -3pt
		,	legend cell align = left
		,	legend style={draw=none}
		,	bar width= 1.75pt
		%,	x tick label as interval=false
		%,	xtickten={-16,-14,-12,...,4}
		%,	yticklabel={}
		,	xtick pos=left
		,	ytick pos=left
		,	xmin=-1
		,	xmax=100
		,	ymin=0.5
		,	ymax=1000
		%,	grid=none
		,	ymajorgrids
		,	log origin=infty
		,	cycle list name=compBankList
		,	ylabel = {Count}
		,	xlabel = {Degree of banks in interbank network}
		,	ylabel near ticks
		,	ylabel style={fill = white, xshift = 1cm}
		]
		\addplot+[,draw=none] table [x=left,y=in] {\DRTX};\addlegendentry{Indegree}
		\addplot+[,draw=none] table [x=left,y=out] {\DRTX};\addlegendentry{Outdegree}
	\end{axis}
\end{tikzpicture}
	\caption{In- and out-degree of banks in the interbank network $B$. Main plot has 60 uniform bins on the interval $[0,360]$, the inset shows the distribution of degrees in the range $[0,100]$.}%
	\label{fig:deg:b:inout}
\end{figure*}
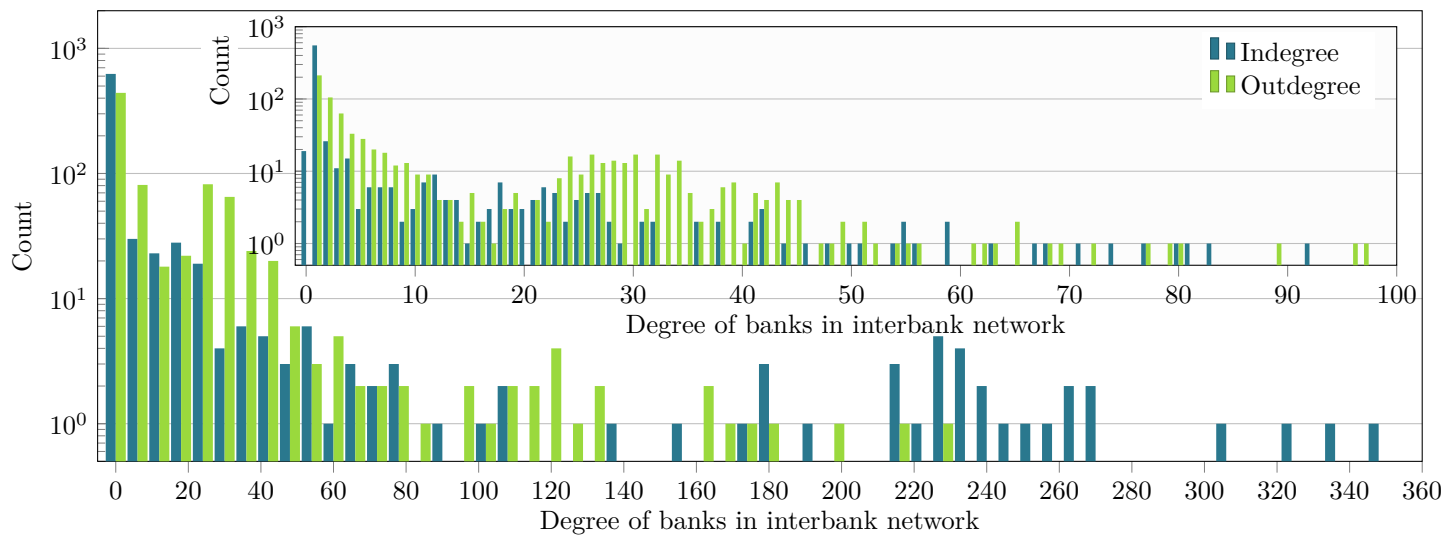
The degree distributions of the banks in the entire liability network and the interbank network are illustrated in \cref{fig:deg:f:inout,fig:deg:b:inout}. 
The in- and out-degree distributions are depicted in \cref{fig:deg:f:inout} for the entire liability network $F$, and in \cref{fig:deg:b:inout} for the interbank network $B$.
In \cref{fig:deg:f:inout,fig:deg:b:inout} the main plots show the whole range, the insets provide a finer resolution in the ranges with higher density.

\begin{table}
	\centering
	\begin{tabular}{lrrcc}
		Network   & Nodes & Links & $\langle C_i \rangle$ & $\langle C_i\rangle^\text{rand}$\\\hline
		Entire network $F$&5,796	&	28,127 & 0.77 & 0.005\\
		Interbank network $B$& 796	&	12,783 & 0.89 & 0.043
	\end{tabular}
	\caption{Number of nodes and links in the entire liability network and in the interbank network. The undirected and unweighted global clustering coefficient $\langle C_i \rangle$ of both networks and the global clustering coefficient of a random graph with same number of nodes and links. The random graph has considerable lower clustering as both networks.}\label{tab:clusterCoeff}
\end{table}
\Cref{tab:clusterCoeff} shows 
the undirected and unweighted global clustering coefficients $\langle C_i \rangle$ of the entire liability network and the interbank network. Clustering coefficients are significantly larger than the clustering coefficients of random graphs with identical number of nodes and vertices $\langle C_i \rangle^\text{rand}$, as shown in \cref{tab:clusterCoeff}.

\begin{figure}
	\centering
		% !TeX root = ../../MSc.tex
\begin{tikzpicture}
	\def\DRT{./data/5000c/hists/CDeg.csv}
	\begin{axis}[
		,	width=0.35\textwidth
		,	height=100pt
		%,	axis on top
		,	scale only axis
		,	ymode=log
		%,	xmode=log
		,	ybar = 0pt% -3pt
		,	legend cell align = left
		,	bar width= 4pt
		,	legend style={draw=none}
		,	xtick pos=left
		,	ytick pos=left
		,	xmin=0
		%,	xmax=0.7
		,	ymin=1
		,	ymax=4000
		%,	grid=none
		,	ymajorgrids
		,	log origin=infty
		,	cycle list name=compBankList
		,	ylabel = {Count}
		,	xlabel = {Degree of firms in the entire network}
		]
		\addplot+[,draw=none] table [x=left,y=in] {\DRT};\addlegendentry{Indegree}
		\addplot+[,draw=none] table [x=left,y=out] {\DRT};\addlegendentry{Outdegree}
	\end{axis}
\end{tikzpicture}  
		\caption{In- and out-degree of firms in the entire liability network. Values are identical for all firms because loans and deposits ($BC$ and $CB$ network, see \cref{sec:construction}) use the same adjacency matrix, based on the bank connections in the commercial register.}\label{fig:deg:companies}
\end{figure}
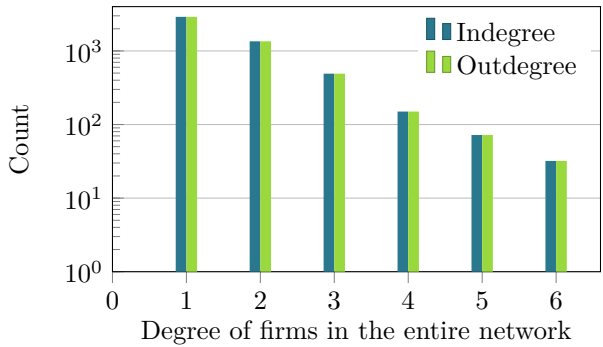
\Cref{fig:deg:companies} shows the degree distribution of firms in the entire liability network (similar to \cref{fig:bankhist}. The degree distribution is restricted to firms with degree $>0$ and contains the 5,000 firms with the highest liabilities in 2008. Note that the in- and out-degree for firms are identical, as the bank connections provided to the commercial register were used for both types of connections between firms and banks, deposits and liabilities.

\section{Systemically important firms and banks in Austria} \label{sec:results}
\begin{figure}
	\centering
	%\fbox
	\begin{tikzpicture}
		\draw[draw=white] (-0.2,-1.2) rectangle(8.8,0.5);
		\begin{axis}[hide axis,scale only axis,height=0pt,width=0pt,colormap name=debtrank,colorbar,point meta min=0,point meta max=0.7,colorbar style={height=180pt,ytick={0,0.1,...,0.7,0.7}}]
			\addplot [draw=none] coordinates {(0,0)};
		\end{axis}
		\node[anchor=north east] (img) at(0,-0.45) {\includegraphics[width=180pt]{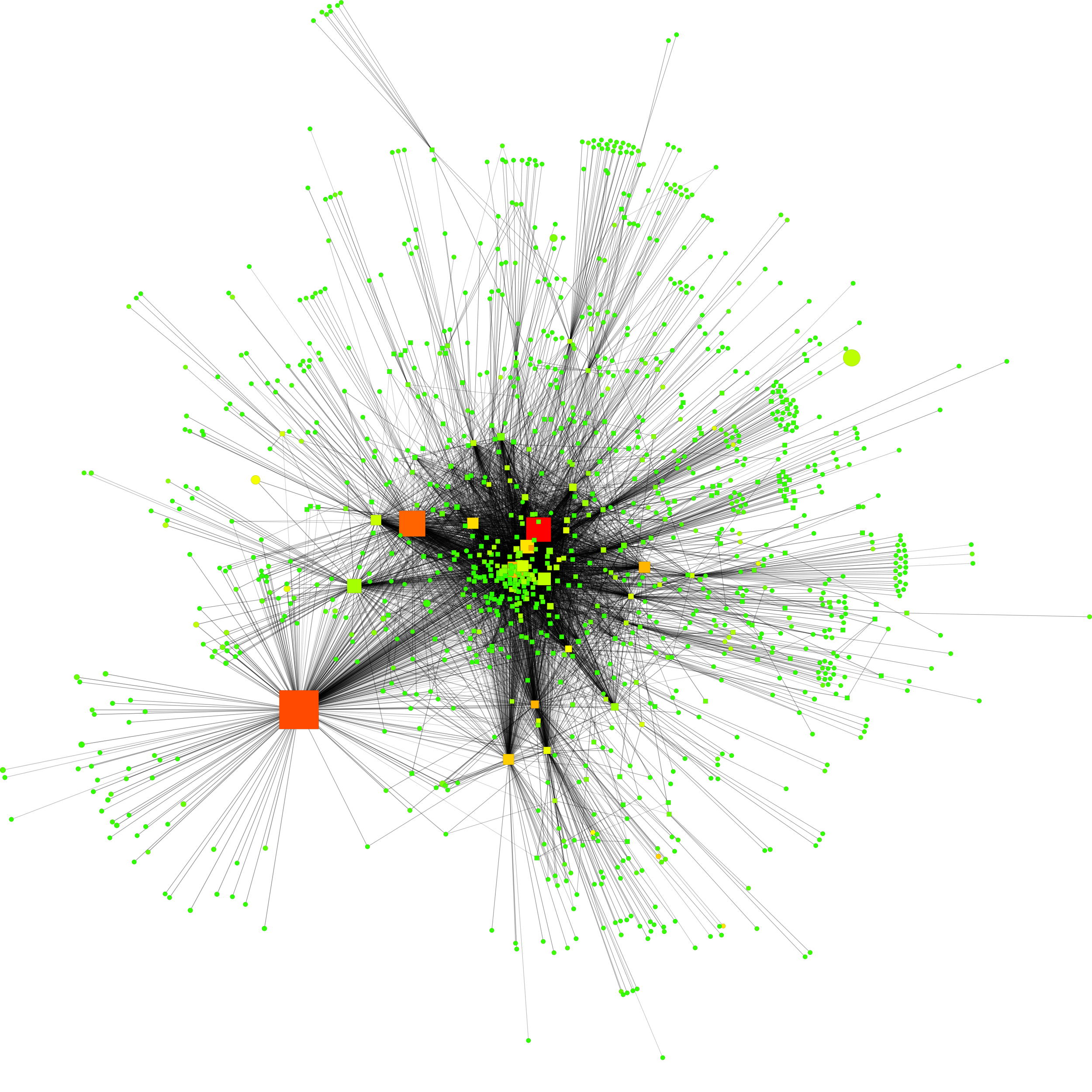}};
		\node[anchor=south east, rotate = 90] (label) at(0.25,-0.25) {DebtRank $R^F$};
	\end{tikzpicture}
	\caption{Subgraph of \cref{fig:graph:50k} with nodes having a DebtRank $R^F~\geq~0.01$. Bank nodes are represented by squares and firms by circles. Node size corresponds to the total assets held by each node. Nodes are colored according to their DebtRank. Nodes representing firms and banks with more assets tend to have a higher DebtRank, but there are also several nodes with a high DebtRank that represent medium-sized banks and firms.}%
	\label{fig:graph:5k:debtrank}
\end{figure}
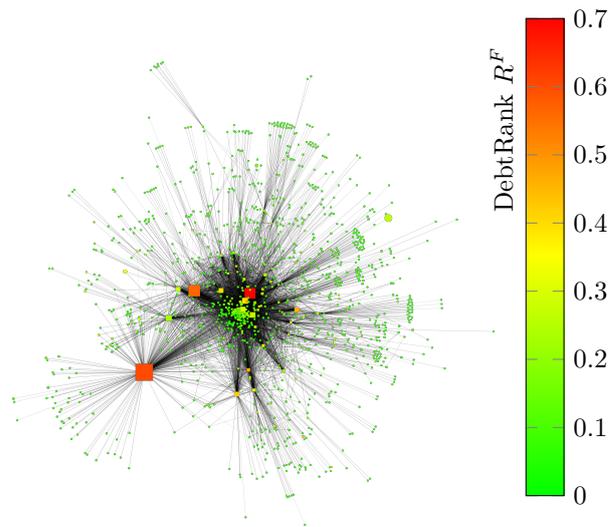
To identify systemically important firms and banks we use DebtRank. DebtRank was originally suggested as a recursive method to determine the systemic relevance of nodes within financial networks \citep{Battiston:2012aa}. It is a quantity that measures the fraction of the total economic value $V$ in the network that is potentially affected by the distress of an individual node $i$, or by a set of nodes $S$. For details see \cref{sec:DebtRank}.

\Cref{fig:graph:5k:debtrank} shows all banks (squares) and firms (circles) with a DebtRank $R^F \geq 0.01$. Node size represents the total assets, whereas the color encodes the DebtRank. Nodes with the highest DebtRanks are typically large banks with substantial total assets, but there are also several nodes with a high DebtRank that represent medium-sized banks and firms. In particular, some medium-sized banks and firms with total assets below 1 bln. EUR exist that have a rather high DebtRank ($\approx 0.4$).

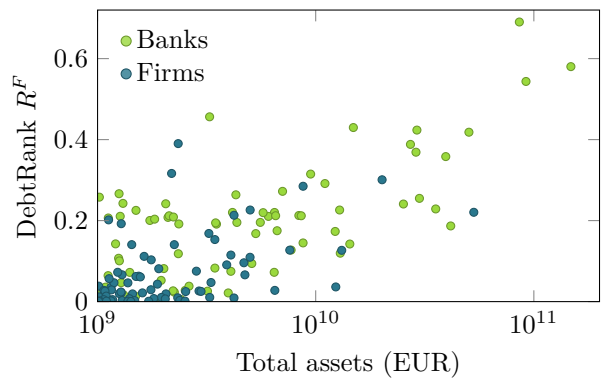
\begin{figure}
	\newcommand{\defZoom}[4]{\def\XMIN{#1}\def\XMAX{#2}\def\YMIN{#3}\def\YMAX{#4}}
	\centering
	\defZoom{1E9}{2E11}{0.0}{0.72}	
	\begin{subfigure}[t]{0.450\textwidth}
		\centering
		% !TeX root = ../MSc.tex
%\pgfplotstableread{./data/BilanzsummeDebtrank.txt}\loadedtable
\def\File{./data/\NWSize/BilanzsummeDebtrankBankCompanies.txt}
\begin{tikzpicture}
\begin{axis}[
,	scale  only  axis
%,	axis y line*=left% the ’*’ avoids arrow heads
,	height=110pt
,	xmode=log
,	width=0.8\textwidth
,	xlabel={Total assets (EUR)}
,	\iftoggle{thesis}{ylabel={}}{ylabel = {DebtRank $R^F$}}
,	ticks=major
%,	xmajorgrids
%,	enlargelimits=false
, 	major grid style={line width=0.1pt,draw=gray!30}
,	xmin = \XMIN
,	xmax = \XMAX
,	ymin= \YMIN
,	ymax=\YMAX
,	legend cell align=left
,	legend pos=north west
%,	cycle list name=compBankList,
,	legend image post style={scale=1.25}
,	clip marker paths=true, axis on top=true
,	ylabel near ticks
,	legend style={fill=white, fill opacity=0.8, draw opacity=1,text opacity=1, draw = none}
]

\addplot[bankColor
	,	only marks
	,	discard if not={Bank}{1}
	,	mark size=1.5pt
	]table[
		x = TotalAssets
	,	y = DebtRank
	,	filter discard warning=false
	]{\File};\addlegendentry{Banks}%\label{plt:scattercost}
\addplot[companyColor
	,	only marks
	,	discard if not={Bank}{0}
	,	mark size=1.5pt
	]table[
		x = TotalAssets
	,	y = DebtRank
	,	filter discard warning=false
	]{\File};\addlegendentry{Firms}%\label{plt:scattercost}
\end{axis}
\end{tikzpicture}
	\end{subfigure}
	\caption{DebtRank of firms and banks plotted against their total assets (as a proxy for firm size) in Euro. Firms with similar DebtRank have differences in their asset sizes of multiple orders of magnitude. Distribution of banks and firms do not seem to be qualitatively different.}
	\label{fig:dr:scatter:balance}
	\centering
\end{figure}
This can also be seen in \cref{fig:dr:scatter:balance}, which shows the DebtRank values of firms and banks plotted against their total assets. In general, firms as well as banks with larger assets tend to have a higher DebtRank. However, firms with similar DebtRank have differences in their total assets of multiple orders of magnitude. The distribution of banks and firms does not seem to be qualitatively different.

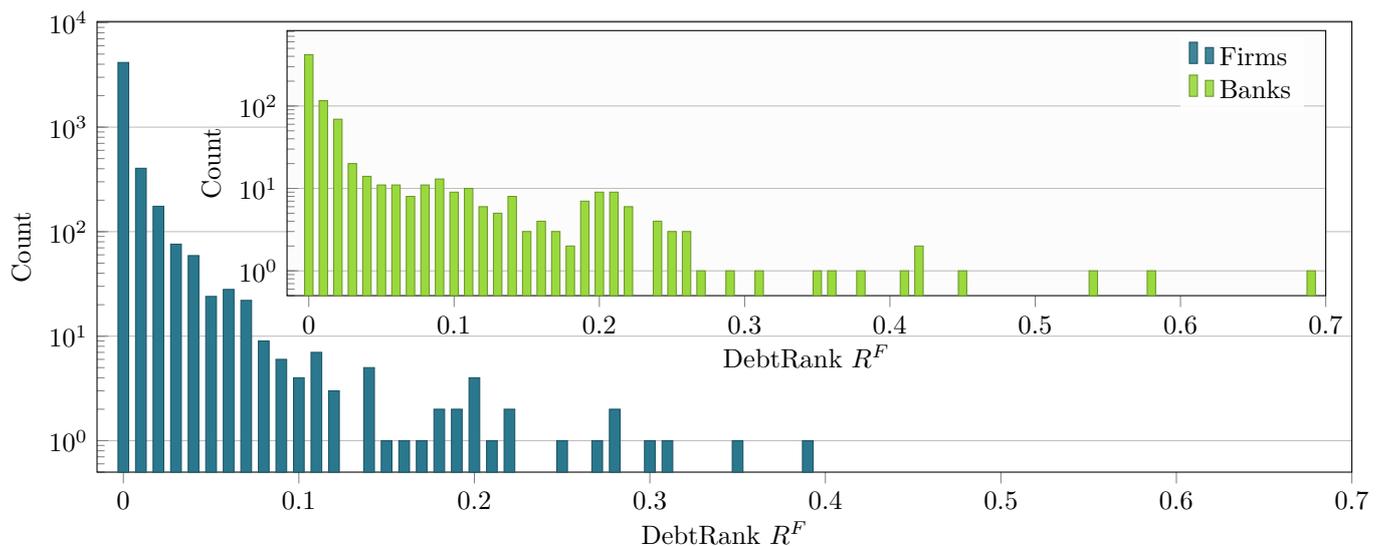
\begin{figure*}
	\centering
	\vspace{15pt}
	% !TeX root = ../../MSc.tex
\begin{tikzpicture}
	\def\DRT{./data/5000c/hists/DR.csv}
	\begin{axis}[
		,	width=0.9\textwidth
		,	height=170pt
		%,	axis on top
		,	scale only axis
		,	ymode=log
		,	ybar
		,	legend cell align = left
		%,	bar width= 3.75pt
		%,	x tick label as interval=false
		,	xtick={0,0.1,0.2,0.3,0.4,0.5,0.6,0.7}
		%,	yticklabel={}
		,	xtick pos=left
		,	ytick pos=left
		,	xmin=-0.015
		,	xmax=0.7
		,	ymin=0.5
		%,	ymax=1
		%,	grid=none
		,	ymajorgrids
		,	log origin=infty
		,	cycle list name=compBankList
		,	ylabel = Count
		,	xlabel = DebtRank $R^F$
		,	ylabel near ticks
		]
		\addplot+[bar width=4pt] table [x=left,y=c] {\DRT};\label{plt:drhist:companies}%\addlegendentry{Companies}
		%				\addplot+[bar width=2.25pt,draw=none] table [x=left,y=b] {\DRT};%\addlegendentry{Banks}
		
	\end{axis}
	\begin{axis}[
		,	width=0.745\textwidth
		,	xshift = 2.5cm
		,	yshift = 2.34cm
		,	axis background/.style={fill=gray!02}
		,	height=100pt
		%,	axis on top
		,	scale only axis
		,	ymode=log
		,	ybar = -3.15pt
		,	legend cell align = left
		,	bar width= 3.15pt
		,	xtick={0,0.1,0.2,0.3,0.4,0.5,0.6,0.7}
		,	xtick pos=left
		,	ytick pos=left
		,	xmin=-0.015
		,	xmax=0.7
		,	ymin=0.5
		%,	ymax=1
		%,	grid=none
		,	ymajorgrids
		,	log origin=infty
		,	cycle list name=compBankList
		,	ylabel = Count
		,	xlabel = DebtRank $R^F$
		,	ylabel near ticks
%		,	ylabel style = {fill = white}
		,	legend style={fill=white, fill opacity=0.9, draw opacity=1,text opacity=1, draw = none}
		]

		% Combined legend for two plots - use refstyle
		\addlegendimage{/pgfplots/refstyle=plt:drhist:companies}\addlegendentry{Firms}
		\addplot[bankColor] table [x=left,y=b] {\DRT};\label{plt:drhist:banks}
		\addlegendimage{/pgfplots/refstyle=plt:drhist:banks}\addlegendentry{Banks}
		
	\end{axis}
\end{tikzpicture}
	\caption{Histogram of DebtRanks $R^F$ in the entire liability network of banks \plotref{plt:drhist:banks} and firms \plotref{plt:drhist:companies}. Banks and firms have a qualitatively similar DebtRank-distribution. The highest DebtRank of a firm is $0.39$.}
	\label{fig:hist:DebtRankFull}
\end{figure*}

	\def\CUTOFF{45}
	\def\BarWidth{8pt}
	\def\HEIGHT{500pt}
	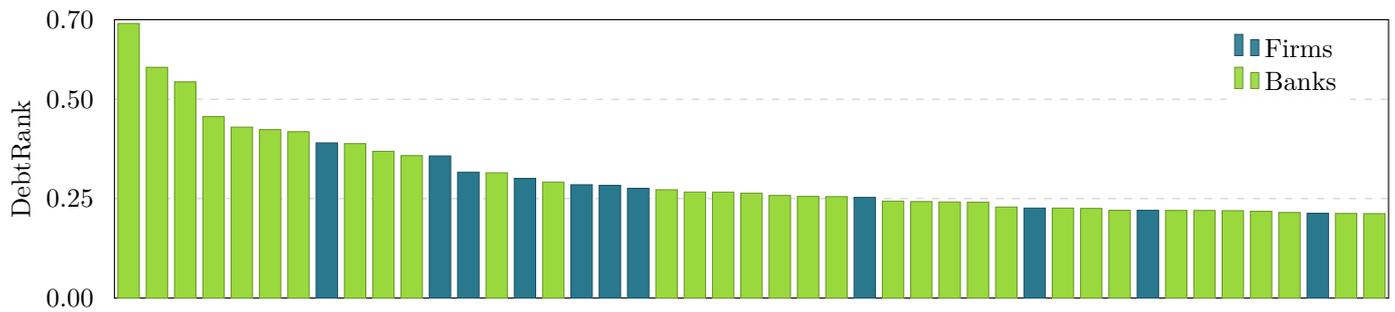
\begin{figure*}
		% !TeX root = ../MSc.tex
\begin{tikzpicture}[]
\def\file{./data/5000c/drSortedBoth.csv}
\pgfplotstableread[col sep=tab,trim cells]{\file}\table
\begin{axis}[
, 	height=150pt
,	width=\textwidth
,	xbar
,	ylabel={DebtRank}
%,	xlabel={Companies/banks ranked by DebtRank}
,	scaled y ticks = false
,	ymin = 0,	ymax =0.7
,	ytick = {0,0.25,0.50,0.7}
,	yticklabels = {0.00,0.25,0.50,0.70}
,	xmin = 0, xmax = \CUTOFF
,	ymajorgrids,
,	xtick = \empty
% ,	axis y line*=left
% ,	axis x line*=bottom
,	bar width= \BarWidth
,	ybar =-\BarWidth
%,	hide y axis
,	major grid style={thin,dashed,black!20}
,	legend pos=north east
,	legend style={fill=white, fill opacity=0.9, draw opacity=1,text opacity=1, draw = none}
,	legend cell align=left
,	every node near coord/.append style={anchor=west,font=\scriptsize}
,	cycle list name=compBankList
,	x tick style={opacity=0}
,	y tick style={opacity=0}
,	xlabel style={yshift=10pt}
]
\addplot+[
    , point meta=explicit symbolic,
    , discard if not={bank}{0},
    , yticklabels from table={\table}{label}
    ] table[
    , trim cells
    , x expr= \coordindex + 0.5
    , y =debtrank
    , meta=label
    ] from \file;
\addlegendentry{Firms}
\addplot+[
    ,   point meta=explicit symbolic
    ,   discard if not={bank}{1}
    ,   yticklabels from table={\table}{label}
    ] table[
    ,   trim cells
    ,   x expr= \coordindex  + 0.5
    ,   y =debtrank
    ] from \file;
\addlegendentry{Banks}
\end{axis}
\end{tikzpicture}
		\caption{DebtRanks of \CUTOFF\ firms and banks sorted by their DebtRank from left to right in decreasing order. With the exception of the three largest banks in Austria (see also \cref{fig:graph:5k:debtrank}), the distribution of banks and firms does not seem to be qualitatively different.}\label{fig:drsorted:mixed}
	\end{figure*}
	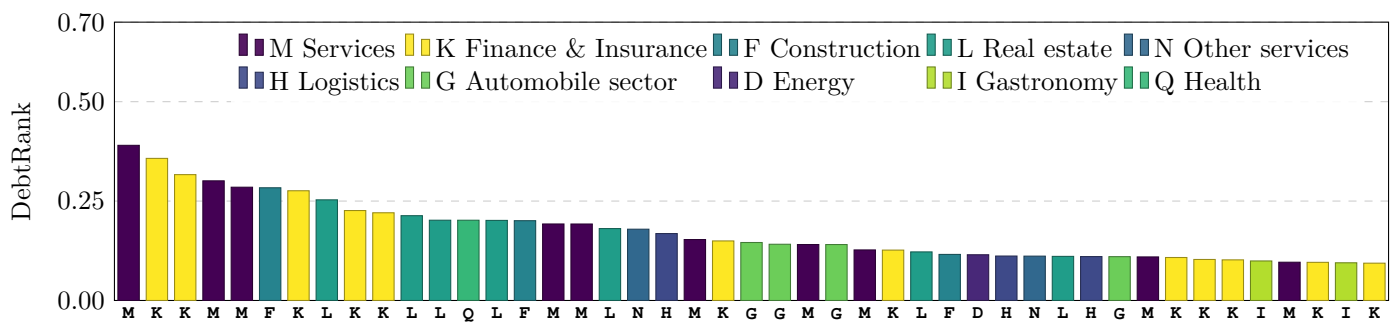
\begin{figure*}
		% !TeX root = ../MSc.tex
\begin{tikzpicture}[]
\def\file{./data/5000c/drSortedComp.csv}
\pgfplotstableread[col sep=tab,trim cells]{\file}\table
\begin{axis}[clip marker paths=true,% axis on top=true
, 	height=150pt
,	width=\textwidth
,	ybar
,	ylabel={DebtRank}
%,	ylabel={Index sorted by DebtRank}
,	scaled y ticks = false
,	ymin = 0,	ymax =0.7
,	ytick = {0,0.25,0.50,0.7}
,	yticklabels = {0.00,0.25,0.50,0.70}
,	xmin = 0, xmax = \CUTOFF % 35 bis Flughafen Wien
,	ymajorgrids,
% ,	axis y line*=left
% ,   axis x line*=bottom
,	bar width= \BarWidth
,	ybar =-\BarWidth
%,	hide y axis
,	major grid style={thin,dashed,black!20}
%,	legend pos=north west
,	legend columns=\iftoggle{thesis}{3}{5}
,	legend style={fill=white, fill opacity=0.9, draw opacity=1,text opacity=1, draw = none}
,	legend cell align=left
,	every node near coord/.append style={anchor=west, rotate=50,font=\scriptsize\ttfamily\bfseries, inner sep = 0.5pt, xshift = 2pt, yshift=3pt, rounded corners=2pt, fill = white}
,	x tick style={opacity=0}
,	y tick style={opacity=0}
,	cycle list name=onaceColors
%,	xtick = \empty
,	xtick = data % add ONACE as xTICK
,	xticklabels from table={\table}{o} % o = 1 ona = 3 digits of ONACE
,	xticklabel style={font=\scriptsize\ttfamily\bfseries, anchor=north,yshift=5pt}
, ]
% M 0 Dienstleistung/Services
% K 1 Finance & 
% F 2 Baugewerbe
% L 3 Immobilien
% Q 9 Health
% N 4 Sonstiges
% H 5 Logistik
% G 6 Kraftfahrzeuge
% D 7 Energy
% I 8 Gastronomy
% S 15 Dienstleistung sonstige
% C 10 Verarbeitendes Gewerbe/Herstellung
% O 11 Government
% E 12 Water & Sewage
% J 13 Kommunikation
% B 14 Mining
% R 16 Art & Entertainment
% A 17 Land & Forst

% this plot ensures that there exists an xtick label at every bar.
% necessary because xtick = data uses only the x values from the first plot
\addplot[forget plot, draw = none
	,	point meta=explicit symbolic,
	,	xtick = data
	,	xticklabels from table={\table}{o}
	]table[
	,	trim cells
	,	x expr= \coordindex +0.5
	,	y =debtrank
	,	meta=o
	] from \file;

	% \pgfplotsset{cycle list shift=-1}

\foreach \i in {0,...,9}{
	\addplot+[
	% ,	nodes near coords
	% ,	nodes near coords align={vertical},
	,	point meta=explicit symbolic,
	,	discard if not={idx}{\i}
%	,	yticklabels from table={\table}{label}
%	,	xtick = data
%	,	xticklabels from table={\table}{o}
	]table[
	,	trim cells
	,	x expr= \coordindex +0.5
	,	y =debtrank
	,	meta=label
	] from \file;
}
% M 0 Dienstleistung/Services
% K 1 Finance & 
% F 2 Baugewerbe
% L 3 Immobilien
% Q 9 Health
% N 4 Sonstiges
% H 5 Logistik
% G 6 Kraftfahrzeuge
% D 7 Energy
% I 8 Gastronomy
\addlegendentry{M Services}
\addlegendentry{K Finance \& Insurance}
\addlegendentry{F Construction}
\addlegendentry{L Real estate}
\addlegendentry{N Other services}
\addlegendentry{H Logistics}
\addlegendentry{G Automobile sector}
\addlegendentry{D Energy}
\addlegendentry{I Gastronomy}
\addlegendentry{Q Health}
\end{axis}
\end{tikzpicture}
	\caption{Firms in different economic sectors ranked by DebtRank (in descending order). The colors correspond to the economic sector in which the firms operate, according to the first level of their OeNACE classification (also used as x-axis label).}\label{fig:drsorted:branch}
\end{figure*}	
In \cref{fig:drsorted:mixed} we show firms and banks in Austria ranked according to their systemic importance measured by DebtRank. \Cref{fig:drsorted:mixed} shows the $45$ banks \plotref{plt:drhist:banks} and firms \plotref{plt:drhist:companies} with the highest DebtRanks, ranked by their DebtRank in decreasing order from left to right. With the exception of the three largest banks in Austria (see also \cref{fig:graph:5k:debtrank}), the distribution of banks and firms does not seem to be qualitatively different. For example, the highest DebtRank of a firm in Austria is $0.39$. Thus, a default of this firm would affect $39\%$ of the Austrian financial system. 

\Cref{fig:drsorted:branch} shows the $45$ firms with the highest DebtRank and additionally provides information about their line of business, according to the first level of their OeNACE code (below the bars), a system to classify economic activities used in Austria~\cite{wirtschaftskammer_osterreich_onace_2008}. Clearly, systemically important firms are found in various industries.

In \cref{fig:hist:DebtRankFull} the distribution of DebtRank values of firms and banks (inset) is depicted with a histogram (70 bins and range $=[0,0.7]$ in both cases). Again, banks and firms have a qualitatively similar DebtRank-distribution. Apparently, systemically important firms contribute systemic risk in a similar way as banks. 

Finally, we estimate the share of systemic risk introduced by firms in the entire liability network. We define $Q_1$ as the ratio of the sum of the DebtRanks of all firms divided by the sum of all DebtRanks in the entire liability network, 
\begin{equation}
Q_1 =  \frac{\sum_{i\in C}R^F_i}{\sum_{i\in F} R^F_i}\ .
\end{equation}
We find $Q_1 = 0.55$ in Austria for 2008. Firms introduce more than half of the systemic risk in the entire liability network network (more than banks). To compare the systemic risk of the interbank network with the systemic risk of the entire liability network we define a similar ratio  
\begin{equation}
Q_2 = \frac{V^B\sum_{i\in B}R^B_i}{V^F\sum_ {i\in F}R^F_i}\,,
\end{equation}
where $V^B$ and $V^F$ refer to the total economic values of the interbank network and the entire liability network, respectively.
In this case we must take the different economic value of the two networks into account, since the DebtRank is a relative measure. 
We find $Q_2= 0.29$ in Austria for 2008, i.e. the total systemic risk of the interbank network amounts to only $29\%$ of the total systemic risk of the entire liability network. 

\section{Conclusions}\label{sec:discussion}
Systemic importance of financial institutions is related to the topology of financial networks to a large extent.  In this work we reconstruct and analyze the to our knowledge largest financial network that has been studied up to now. This financial network consists of 51,980 firms and \nBanks banks representing $80.2\%$ of total liabilities towards banks by firms and all interbank liabilities from the entire Austrian banking system visualized in \cref{fig:graph:50k}.

We find that firms induce systemic risk in a similar way as banks. Banks and firms have a qualitatively similar distribution of systemic importance (see \cref{fig:hist:DebtRankFull}). In particular, we identify several medium-sized banks and firms with total assets below 1 bln. EUR in Austria that are systemically important in the entire financial network. Systemic importance of these firms is primarily driven by their position in the network. Moreover, these firms belong to various industries (\cref{fig:drsorted:branch}). We further find that banks and firms of similar systemic importance have differences in their asset sizes of multiple orders of magnitude (\cref{fig:dr:scatter:balance}). Our main result is that, overall, firms introduce slightly more systemic risk than banks and that in Austria for the year 2008 the total systemic risk of the interbank network amounts to only $29\%$ of the total systemic risk of the entire financial network consisting of firms and banks. 

These results come with three caveats due to partially missing and partly inaccurate data. First of all, the analyzed financial network is reconstructed and not directly taken from empirical data. However, the reconstruction process only involves reconstructing the weights from the (unweighted) adjacency matrix, which is directly taken from empirical data. Moreover, for a large subset of firms, the liabilities towards banks ($42.4\%$ of total liabilities towards banks) are known exactly and do not need to be reconstructed since these firms are only customers of one bank (\cref{fig:bankhist}). The interbank liabilities are also not reconstructed. Second, interbank liabilities from the Austrian banking system are fully anonymized and linearly transformed. Thus, there is a small uncertainty in the absolute value of the interbank liabilities, which also introduces some uncertainty in the matching process of the various datasets and third, our analysis involves only one snapshot of the Austrian financial system in 2008 -- the only year, where the necessary datasets overlap. 

It would be interesting to extend this study to other countries and to investigate the evolution of similar large financial networks that consist of firms and banks and represent interbank liabilities and the liabilities towards banks by firms. It would further be interesting to go beyond the scope of this work and study a financial network that truly represents all liabilities between all economic agents from all sectors (households, non-financial and financial firms and the government sector) in an entire national economy.

Though further investigation might be necessary to confirm our findings for other countries and over longer time horizons, we believe that this contribution is a valuable first comprehensive quantification of the financial interrelationships between the financial and the real economy. We show that the notion of systemically important financial institutions (SIFIs) or global and domestic systemically important banks (G-SIBs or D-SIBs) can be straightforwardly extended to firms. In particular, we identify several medium-sized banks and firms with total assets below 1 bln. EUR in Austria that are systemically important. In conclusion, our analysis suggests that not only systemically important financial institutions but also systemically important firms must be subject to macro-prudential regulation.

\section*{Acknowledgments}
We thank Anita Wanjek and Michael Miess for helping us with the manuscript.

\section*{References}
\bibliographystyle{elsarticle-harv}
\bibliography{references,../../cosy/bibliography/trunk/econophysics}

\cleardoublepage
\appendix

\section{DebtRank}\label{sec:DebtRank}
The financial dependencies of the nodes in the network are given in a liability matrix $L$ with entries $L_{ij}$ denoting that node $j$ has given node $i$ a loan (or investment/deposit) of size $L_{ij}$. 
Additionally, there is a capital (or equity) vector $C$ with entries $C_i$ denoting the capital of node $i$. 

The relative economic value of a node $i$ is given by 
\begin{equation}
	v_i = \frac{L_i}{\sum_j L_j}\label{eq:relEconVal}
\end{equation}
where $L_i = \sum_j L_{ji}$ is the sum of the outstanding liabilities of node $i$.

The default of node $i$ then affects all nodes $j$ where $L_{ij}> 0$.
The impact of the default of $i$ on $j$ is defined as
\begin{equation}
	W_{ij} = \min\left(\frac{L_{ij}}{C_j},1\right)\ .
\end{equation}
The impact of a shock is thus measured as the fraction of capital loss due to the credit default. It is therefore a value in the range $[0,1]$ with the semantics $W_{ij} = 0$ if the default of node $i$ does not affect node $j$, and $W_{ij} = 1$ if the default of node $i$ results in a loss that matches or exceeds the capital of node $j$. 

The economic value of the impact is obtained by multiplying the impact with the relative economic value from \cref{eq:relEconVal}. The economic value of the impact of $i$ on its neighbors is therefore given by
\begin{equation}
	I_i =  \sum_j W_{ij}v_j\label{eq:impactV1}
\end{equation}

Though if the neighbors of $i$ do not have enough capital to compensate for the default of $i$, they default themselves, leading to an impact on their neighbors and reverberations in the network along paths in the impact network $W$.
To prevent infinite cycles, Battiston et al. proposed to only take walks without repeating links.
This is achieved by introducing two additional state variables for each node, $s_i$ and $h_i$ (both dependent on the time step $t$). 
The variable $s_i$ takes one of three values:
\begin{center}
	\centering
	\begin{tabular}{ccl}
		$s_i(t)$  & & Interpretation\\\hline
		U   &    &   Node $i$ is undistressed at time $t$\\
		D   &    &   Node $i$ is in distress at time $t$\\
		I   &    &   Node $i$ is inactive at time $t$
	\end{tabular}
\end{center}
The variable $h_i$ has a value in the range $[0,1]$ and is the level of distress, with $0$ meaning \emph{undistressed} and $h_i(t) = 1$ in the case of default. 
The value of $h_i(t)$ is defined as
\begin{align}
	h_i(t) &= \min\left(1, h_i(t-1) + \sum_{j|s_j(t-1)=D}W_{ji}h_j(t_1)\right)\ ,\label{eq:drH}
	\intertext{whereas $s_i(t)$ is given by:}
	s_i(t) &=
	\begin{cases}
		D  &  \text{if $h_i(t) > 0 \land s_i(t-1)\neq I$},\\
		I  &  \text{if $s_i(t-1) = D$},\\
		s_i(t-1) &  \text{otherwise}
	\end{cases}\label{eq:drS}
\end{align}

To calculate the DebtRank of a node $d$ ($d$ for \emph{defaulting}), the distress $h_i$ and status $s_i$ at time step $t=1$ are initialized as follows:
\begin{align}
	h_i(1) &= \begin{cases}
		1 & \text{if $i = d$}\\
		0 & \text{otherwise}
	\end{cases}
	\\
	s_i(1) &= \begin{cases}
		D & \text{if $i = d$}\\
		U & \text{otherwise.}
	\end{cases}\ .\label{eq:drInit}
\end{align}
Then the values of $s_i$ and $h_i$ are calculated for every node $i$ and time step $t$, according to \cref{eq:drH,eq:drS} until all nodes are either inactive or undistressed at $t=T$.
The DebtRank of node $d$ can then be calculated as the sum of the distress in the whole network at time $t=T$ reduced by the distress at the beginning, i.e. the initial distress of node $d$ at $t=1$:
\begin{equation}
	R_d = \sum_i h_i(T)v_i - h_d(1)v_d .\label{eq:drReduced}\
\end{equation}

It would be possible to calculate the combined DebtRank of a set $S$ of simultaneously defaulting nodes by replacing the $i=d$ conditions in the initialization (\cref{eq:drInit}) by $i\in S$ and changing \cref{eq:drReduced} to one of the following equations:
\begin{align}
	R_S &= \sum_i h_i(T)v_i - \sum_{d\in S}h_d(1)v_d \label{eq:drSetRed}
	\\
	R_S &= \sum_i h_i(T)v_i\label{eq:drSet}\ .
\end{align}
\Cref{eq:drSetRed} excludes the impact of the initial shock, whereas \cref{eq:drSet} does not. 

In this work, DebtRank is calculated on two different networks, the interbank network $B$ and the entire liability network $F$. We use $R^F$ and $R^B$ to discern between the two measures.

\end{document}